\documentclass[aps,prl,twocolumn,showpacs,amsmath,amsmath,amssymb,amsfonts]{revtex4-1}
\usepackage{graphicx}
\usepackage{dcolumn}
\usepackage{bm}
\usepackage{color}
\usepackage{multirow}

\begin{document}
\author{Weibin Li,$^{1,3}$ Daniel Viscor,$^1$ Sebastian Hofferberth,$^2$ and Igor Lesanovsky$^1$}
\affiliation{$^1$School of Physics and Astronomy, The University of Nottingham, Nottingham, NG7 2RD, United Kingdom}
\affiliation{$^2$5. Physikalisches Institut, Universit\"{a}t Stuttgart,Pfaffenwaldring 57, 70569 Stuttgart, Germany}
\affiliation{$^3$School of Physics, Huazhong University of Science and Technology, Wuhan 430074, China}
\title{Electromagnetically induced transparency in an entangled medium}
\date{\today}
\begin{abstract}
We theoretically investigate light propagation and electromagnetically induced transparency (EIT) in a quasi one-dimensional gas in which atoms interact strongly via exchange interactions. We focus on the case in which the gas is initially prepared in a many-body state that contains a single excitation and conduct a detailed study of the absorptive and dispersive properties of such a medium. This scenario is achieved in interacting gases of Rydberg atoms with two relevant $S$-states that are coupled through exchange. Of particular interest is the case in which the medium is prepared in an entangled spinwave state. This, in conjunction with the exchange interaction, gives rise to a non-local susceptibilty which --- in comparison to conventional Rydberg EIT --- qualitatively alters the absorption and propagation of weak probe light, leading to non-local propagation and enhanced absorption. \end{abstract}


\maketitle

Currently there is a great interest in studying the phenomenon of electromagnetically induced transparency (EIT)~\cite{fleischhauer_EIT_2005} in gases of interacting Rydberg
atoms~\cite{saffman_quantum_2010,pritchard_cooperative_2010,friedler_long-range_2005,ates_eit_2011,olmos_2011,gunter_2012,dudin_strongly_2012,yan_2012,garttner_2013}.
The strong interaction between atoms excited to high-lying Rydberg
states translates to an effective interaction between
photons~\cite{gorshkov_photon-photon_2011,peyronel_2012,gorshkov_dissipative_2013}.
This, on the one hand, is of great technological importance as it allows the realization of photon
switches~\cite{baur_single_2014} and photon-photon phase
gates~\cite{friedler_long-range_2005,gorshkov_photon-photon_2011} for quantum information
processing. On the other hand it permits to probe novel quantum
phenomena such as bound states of photons~\cite{firstenberg_attractive_2013,Bienias14}, the crystallization of
photons~\cite{otterbach_wigner_2013} and non-linear and non-local light-matter
interaction~\cite{sevincli_nonlocal_2011,petrosyan_electromagnetically_2011}. The key ingredient underlying this physics is an essentially classical
density-density interaction among Rydberg
states~\cite{pohl_dynamical_2009,lesanovsky_many-body_2011,schaus_observation_2012}.
In certain regimes these systems can however exhibit resonant dipole-dipole
interactions~\cite{anderson_resonant_1998,Ditzhuijzen08,nipper_highly_2012}
which then effectuate excitation exchange among different Rydberg states that can establish coherence between distant atoms, thereby leading to a different kind of non-locality.

\begin{figure}
\centering
\includegraphics*[width=0.85\columnwidth]{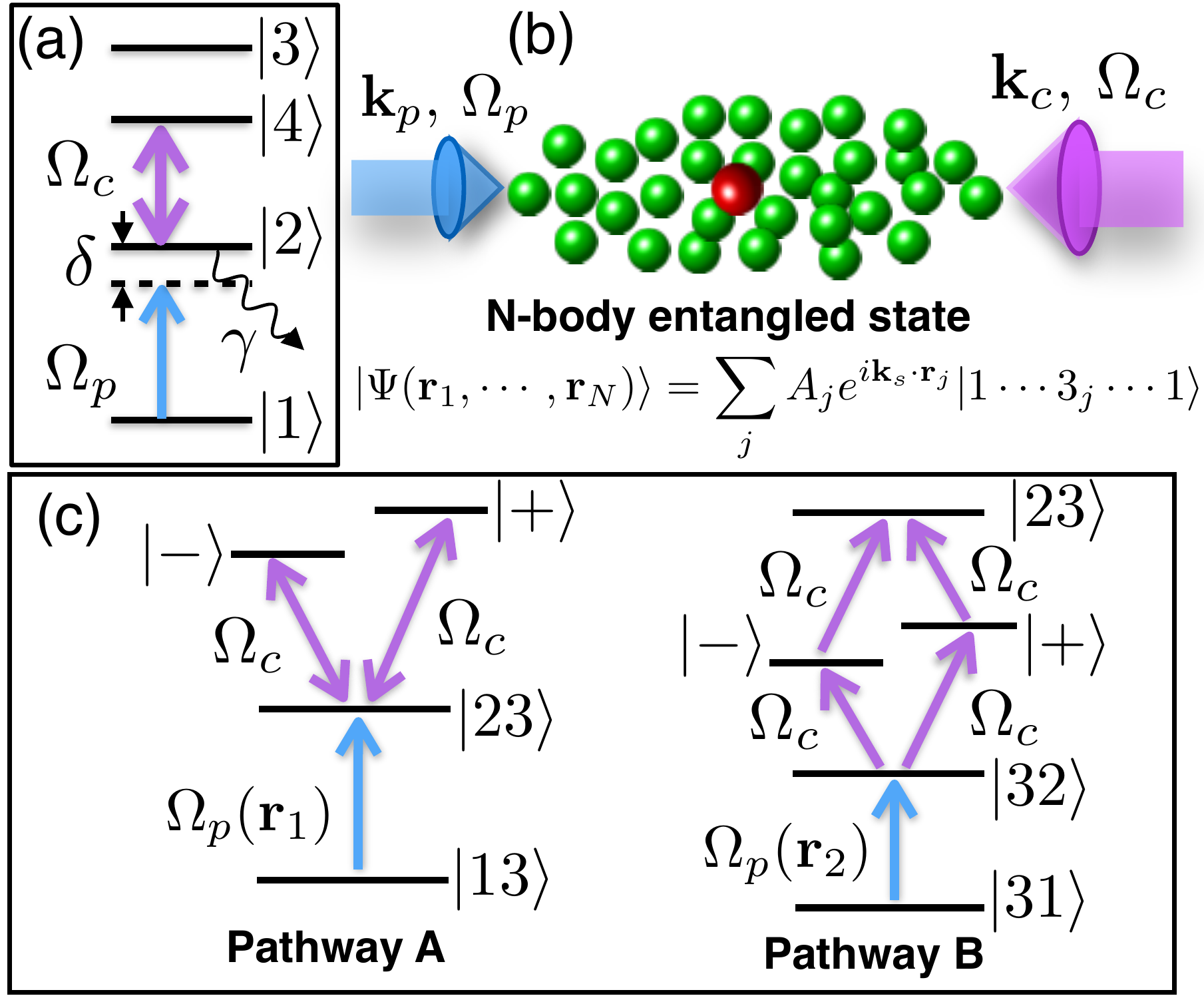}
\caption{(a) We consider atoms with four relevant states, with $|1\rangle$, $|2\rangle$ and $|4\rangle$ forming an EIT ladder system, where the transitions $|2\rangle\leftrightarrow|4\rangle$ and $|1\rangle\leftrightarrow|2\rangle$ are coupled by a classical field $\Omega_c$ and single photom field $\Omega_p$, respectively.
(b) Initially the atomic gas is prepared in an entangled many-body state $|\Psi(\mathbf{r}_1,\cdots, \mathbf{r}_N)\rangle$, in which a single excitation in the Rydberg state $|3\rangle$ is coherently shared by the entire ensemble. (c) Possible pathways for exciting the two atom state $|23\rangle$: pathway A is local (starting from $|13\rangle$) while pathway B is non-local in the sense that $|32\rangle$ appears as intermediate state. The two pathways A and B result in local and non-local susceptibility, respectively.}
\label{fig:system}
\end{figure}
In this work we consider an atomic medium that is initially prepared in a many-body state which contains a single Rydberg excitation and investigate EIT in a setting with two relevant Rydberg states where excitations can be exchanged between distant atoms \cite{gunter_observing_2013,Ditzhuijzen08}.
When the initial state is a spinwave~\cite{dudin_strongly_2012,dudin_emergence_2012} the \emph{linear} susceptibility of the medium acquires a non-local term, in contrast to current studies using single Rydberg states~\cite{sevincli_nonlocal_2011,petrosyan_electromagnetically_2011}. We demonstrate that the exchange interation furthermore leads to new features: it can produce a super-exponential extinction of weak probe light as well as an altered ``non-local'' propagation that both result from the fact that the photon field at one point of the medium can act as a source at a second (distant) position. We discuss the underlying mechanism --- namely the establishment of quantum coherence through exchange --- within a two atom model and subsequently conduct a numerical study of a one-dimensonal atomic gas. We conclude by discussing a simplified analytical model that illustrates the new absorption features and in particular the interplay of the phases of the involved light fields and the initial spinwave. Beyond being of fundamental interest our results indicate a route towards probing excitonic dynamics in interacting atomic gases~\cite{maxwell_storage_2013,bettelli_exciton_2013} and ways for enhancing the absorptive properties of Rydberg gases which is e.g. important for implementing efficient photon switches \cite{baur_single_2014}.

\indent{\it The system, initial state and relevant interactions.---} We consider a gas of $N$ atoms with positions $\textbf{r}_k$ and a level structure that is depicted in Fig.~\ref{fig:system}a: $|1\rangle$ and $|2\rangle$ are ground- and low-lying excited states, respectively, and the two upper levels $|3\rangle$ and $|4\rangle$ are Rydberg $S$-states.
The states $|1\rangle,\,|2\rangle$ and $|4\rangle$ are driven by two light fields in an EIT configuration: the $|1\rangle\leftrightarrow|2\rangle$-transition is weakly coupled by a single-photon (probe) field with detuning $\delta$, central wave vector $\mathbf{k}_p$ and Rabi frequency $\Omega_{p}(\mathbf{r},t)=\mu_{21}\mathcal{E}(\mathbf{r},t)\exp[i(\mathbf{k}_p\cdot\mathbf{r})]$. Here $\mu_{21}$ is the transition dipole moment and $\mathcal{E}(\mathbf{r},t)$ is the slowly varying envelope. The $|2\rangle\leftrightarrow |4\rangle$-transition is resonantly coupled by a strong classical laser field with wave vector $\mathbf{k}_c$ and Rabi frequency $\Omega_{c}(\mathbf{r})=\Omega\exp[i(\mathbf{k}_c\cdot\mathbf{r})]$. Within the standard rotating wave approximation the corresponding single particle Hamiltonian is then given by $\hat{H}_j=-\delta(\hat{\sigma}_{22}^{j}+\hat{\sigma}_{44}^{j})
-(1/2)[\Omega_p(\mathbf{r}_j)\hat{\sigma}_{21}^{j}+\Omega_c(\mathbf{r}_j)\hat{\sigma}_{42}^{j}
+\text{h.c.}]$ with $\hat{\sigma}_{ab}^{j}=|a\rangle_j\!\langle b|$ \cite{fleischhauer_EIT_2005}. In this description the probe field strength $\mathcal{E}(\mathbf{r})$ appears as a classical field. For a single-photon pulse, this this is equivalent to a field theoretical treatment with $\mathcal{E}(\mathbf{r})$ being the single-photon probability amplitude~\cite{fleischhauer_EIT_2005}.

The central element of this study is the initial state of the atomic gas: it contains a single coherently delocalized excited atom in level $|3\rangle$ with all other atoms being in the ground state (see Fig.~\ref{fig:system}b), i.e., a spinwave or super-atom state $|\Psi(\mathbf{r}_1,\cdots, \mathbf{r}_N)\rangle =\sum_jA_je^{i\mathbf{k}_s\cdot
\mathbf{r}_j}|1\cdots 3_j\cdots 1\rangle$. The probability amplitudes $A_j$ and the spinwave vector $\mathbf{k}_s$ are determined by the preparation
process, e.g., via a single-photon pulse or the use of stronger pulses and exploiting the dipole blockade~\cite{dudin_observation_2012,bariani_dephasing_2012,dudin_emergence_2012,petrosyan_2013}.
Owed to this initial state and due to the fact that we are only considering single-photon probe field pulses (on the $|1\rangle\leftrightarrow|2\rangle$-transition) there will always be at most one atom in state $|3\rangle$ and one in state $|4\rangle$. Therefore only two interactions are relevant: the van-der-Waals interaction, $\hat{V}^{d}_{jk}= V^d_{jk}\, \hat{\sigma}_{33}^{j}\, \hat{\sigma}_{44}^{k}$, and the exchange interaction \cite{walker_consequences_2008}, $\hat{V}^{e}_{jk}= V^e_{jk}\,[\hat{\sigma}_{34}^{j}\,
\hat{\sigma}_{43}^{k}+\mathrm{h.c.}]$, between an atom in state $|3\rangle$ (at position $\textbf{r}_j$) and a second atom in state $|4\rangle$ (at position $\textbf{r}_k$). We consider two Rydberg $S$-states for which the corresponding interaction strengths can be parameterized as
$V^{d,e}_{jk}=C_{d,e}/|\mathbf{r}_{j}-\mathbf{r}_{k}|^6$ where $C_{d,e}$ are dispersion coefficients \cite{rick_thesis}. This leads us to the following Master equation governing the evolution of the density matrix $\rho$ of the system: $\dot{\rho}=-i[\hat{H},\rho]+\gamma\sum_j(\hat{\sigma}_{12}^{j}\rho
\hat{\sigma}_{21}^{j}-\{\hat{\sigma}_{22}^{j},\rho\}/2)$, where $\hat{H}=\sum_j
\hat{H}_j + \sum_{jk}[\hat{V}^{d}_{jk}+\hat{V}^{e}_{jk}]/2$ is the total
Hamiltonian. Here, $\gamma$ stands for the decay rate from state $|2\rangle$ while the decay from the Rydberg states $|3\rangle$ and $|4\rangle$ is neglected, which is justified on sufficiently short time scales due to a typical Rydberg state lifetime on the order of $100$ $\mu$s.

\indent{\it Optical response of two atoms.---} In order to get an understanding of the underlying physical mechanisms we consider first an ensemble of two atoms with the initial state
$|\Psi(\mathbf{r}_1,\mathbf{r}_2)\rangle=[|13\rangle +
e^{i\mathbf{k}_s\cdot\mathbf{r}_{12}}|31\rangle]/\sqrt{2}$, where $\mathbf{r}_{12}\equiv\mathbf{r}_{1}-\mathbf{r}_{2}$. The optical response is given by the coherence between the single atom states $|1\rangle$ and $|2\rangle$. For the atom positioned at $\mathbf{r}_1$ this is $\sigma(\mathbf{r}_1)=\text{Tr}\rho\hat{\sigma}_{21}^1$. To calculate this quantity we compute the stationary state \cite{meta-stable} of the Master equation perturbatively \cite{boyd_nonlinear_2008,fleischhauer_EIT_2005} to first order in the probe field Rabi frequency $|\Omega_p(\mathbf{r}_{1,2})|$ ($|\Omega_p(\mathbf{r})|\ll |\Omega|\,, \gamma$) using the state $|\Psi(\mathbf{r}_1,\mathbf{r}_2)\rangle$ as initial condition. This yields
\begin{flalign}
\label{eq:atomcoh}
\sigma(\mathbf{r}_1)=-\frac{\mu_{21}}{2}[\alpha^+_{\mathbf{r}_1\mathbf{r}_2}\mathcal{E}(\mathbf{r}_1)+\alpha^-_{\mathbf{r}_1\mathbf{r}_2}e^{i\Phi_{\mathbf{r}_1\mathbf{r}_2}}\mathcal{E}(\mathbf{r}_2)],
\end{flalign}
where $\alpha^{\pm}_{\mathbf{r}_1\mathbf{r}_2}=\frac{\delta-V^+_{12}}{\Omega^2-2i\Gamma
[\delta-V^+_{12}]}\pm\frac{\delta-V^-_{12}}{\Omega^2
-2i\Gamma[\delta-V^-_{12}]}$ with $\Gamma=\gamma-2i\delta$. The energies
$V^{\pm}_{12}=V^d_{12}\pm V^e_{12}$ belong to the delocalized
excitonic Rydberg pair states $|\pm\rangle=(|34\rangle\pm|43\rangle)/\sqrt{2}$ which are eigenstates
of the interaction Hamiltonian $\hat{V}^{d}_{12}+\hat{V}^{e}_{12}$.
Furthermore, we have abbreviated by
$\Phi_{\mathbf{r}_1\mathbf{r}_2}=\mathbf{K}\cdot\mathbf{r}_{12}$ the relative phase difference of the probe field which depends on the relative wavevector $\mathbf{K}=\mathbf{k}_s-\mathbf{k}_c-\mathbf{k}_p$ defining a phase matching relation~\cite{boyd_nonlinear_2008}.

The unusual property of the coherence $\sigma(\mathbf{r}_1)$ with respect to conventional EIT is that it depends not only on the local probe field strength $\mathcal{E}(\mathbf{r}_1)$, but also on the strength of the probe field \emph{at the location of the second atom} $\mathcal{E}(\mathbf{r}_2)$. This is the key finding of this work. Note, that this non-local optical response features a linear dependence on the probe field, which differs from the one encountered in former studies of Rydberg EIT, where the non-locality is typically generated in higher orders~\cite{sevincli_nonlocal_2011,petrosyan_electromagnetically_2011}.

The emergence of the non-local term is a direct consequence of the interplay between the exchange interaction and the initial entangled state. The role of the exchange becomes apparent in Eq. (\ref{eq:atomcoh}): for vanishing exchange, i.e., $V^e_{12}=0$, we find $\alpha^-_{\mathbf{r}_1\mathbf{r}_2}=0$ and hence the non-local term of $\sigma(\mathbf{r}_1)$ vanishes as well. To see how the non-local coherence is established when $V^e_{12}\neq 0$ and to understand the role of the initial state it is instructive to analyze the two excitation pathways of the Rydberg pair state $|23\rangle$ that are sketched in Fig.~\ref{fig:system}c. The left panel shows the conventional local pathway in which the probe field $\Omega_p(\mathbf{r}_1)$ establishes a coherence between the states $|13\rangle$ and $|23\rangle$ whose precise value is determined by the coupling of state $|23\rangle$ to the excitonic states $|\pm\rangle$. The panel on the right shows the non-local pathway for the excitation of state $|23\rangle$ starting from $|31\rangle$. This can establish a coherence in the first atom provided that the initial state contains a $|31\rangle$-component, which is the case for $|\Psi(\mathbf{r}_1,\mathbf{r}_2)\rangle$. Note that this requires the excitonic states $|+\rangle$ and $|-\rangle$ to be non-degenerate, i.e. $V^e_{12}\neq 0$, otherwise the two paths from $|32\rangle$ to $|23\rangle$ interfere destructively.

\begin{figure}
\centering
\includegraphics*[width=0.93\columnwidth]{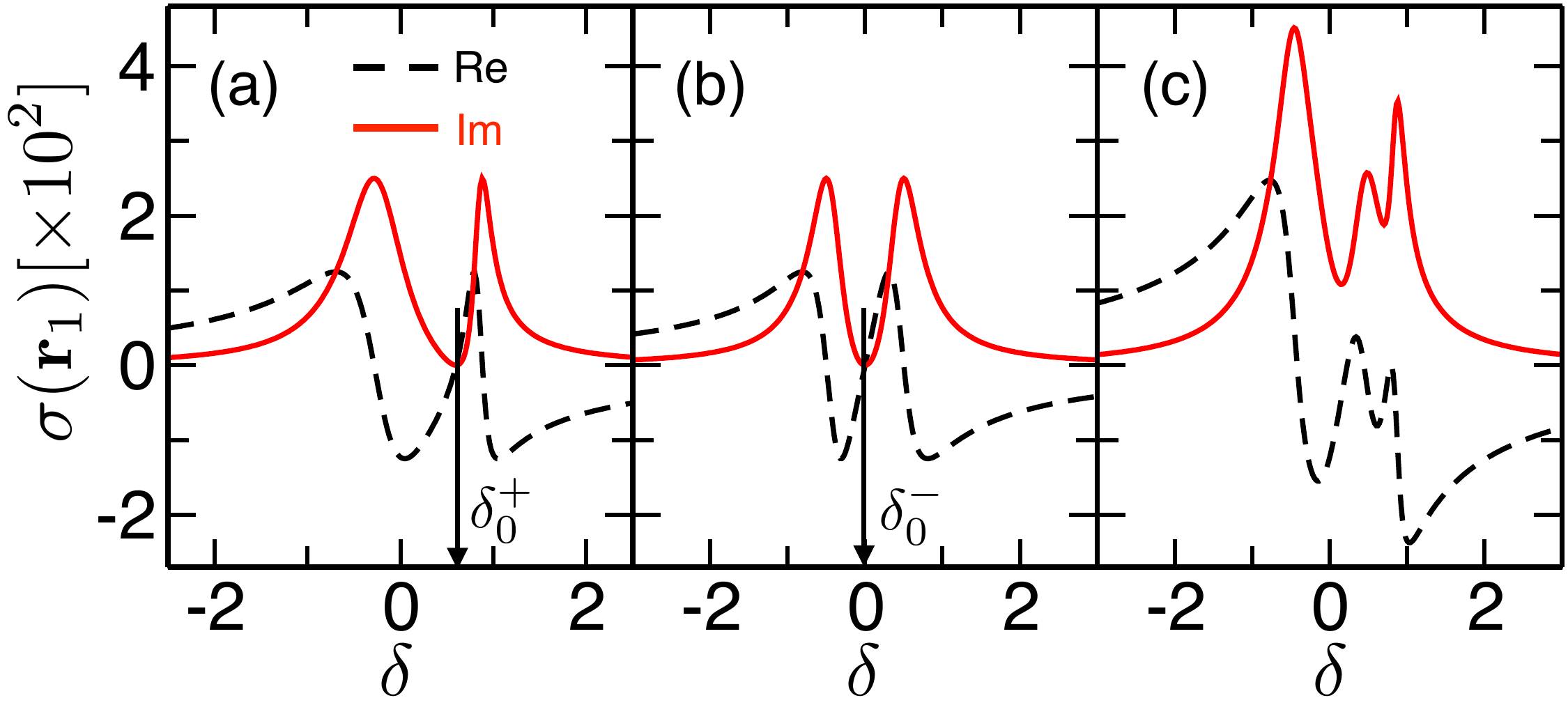}
\caption{Real (dashed) and imaginary (solid) part of the coherence
	$\sigma(\mathbf{r}_1)$ for different initial states. The parameters are
	$\Omega=1.0\gamma$, $V^d_{12}=V^e_{12}=0.3\gamma$, and $\mu_{21}\mathcal{E}(\mathbf{r}_{1,2})=0.05\gamma$. (a) Here $|\Psi(\mathbf{r}_1,\mathbf{r}_2)\rangle=[|13\rangle +
|31\rangle]/\sqrt{2}$ which is coupled solely to the state $|+\rangle$. (b) Here $|\Psi(\mathbf{r}_1,\mathbf{r}_2)\rangle=[|13\rangle -
|31\rangle]/\sqrt{2}$ which is coupled to the anti-symmetric exciton state $|-\rangle$.  (c) Here the initial state contains a localized excitation $|\Psi(\mathbf{r}_1,\mathbf{r}_2)\rangle=|13\rangle$ which can couple to both excitonic states and gives rise to a more intricate behavior of the coherence.}
\label{fig:susceptibility}
\end{figure}
We conclude the discussion of two atoms by analysing the role of the phase in the initial state $|\Psi(\mathbf{r}_1,\mathbf{r}_2)\rangle$. To simplify the picture we assume $C_d=C_e$ and that the projection of  $\mathbf{k}_p+\mathbf{k}_c$ on the vector $\mathbf{r}_{12}$ is zero such that $\Phi_{\mathbf{r}_1\mathbf{r}_2}=\mathbf{k}_s\cdot\mathbf{r}_{12}$. Of particular relevance are the two extreme cases in which $\exp(i\Phi_{\mathbf{r}_1\mathbf{r}_2})=\pm 1$. For a spatially homogeneous probe field, $\Omega_p(\mathbf{r}_1)=\Omega_p(\mathbf{r}_2)$, the initial state can only couple to one of the excitonic pairs states $|\pm\rangle$ due to conservation of parity. This becomes directly visible in the real and imaginary part of the coherence which we show in Fig.~\ref{fig:susceptibility} as a function of the detuning $\delta$ for different initial states. For $\exp(i\Phi_{\mathbf{r}_1\mathbf{r}_2})=1$ (Fig.~\ref{fig:susceptibility}a) we find an asymmetric EIT feature with a vanishing coherence at the detuning $\delta^+_0=V^+_{12}$ that corresponds to the interaction energy of the symmetric exciton state. In the case $\exp(i\Phi_{\mathbf{r}_1\mathbf{r}_2})=-1$ (Fig.~\ref{fig:susceptibility}b) we find a symmetric EIT feature centered at $\delta^-_0=V^-_{12}$, the energy of the anti-symmetric exciton state. Note, that $\delta^-_0=0$ due to the choice of equal strengths for the exchange and van-der-Waals interaction in this particular example. For completeness we show in Fig.~\ref{fig:susceptibility}c the coherence for an initial state in which solely the atom at position $\mathbf{r}_2$ is excited. In this localized case the initial state has no defined parity and hence couples to both excitonic states. The coherence of the first atom is then given by $\sigma_{\text{loc}}(\mathbf{r}_1)=\sigma_+(\mathbf{r}_1)+\sigma_-(\mathbf{r}_1)$ which displays an intricate behavior with multiple peaks.

\begin{figure}
\centering
\includegraphics*[width=0.93\columnwidth]{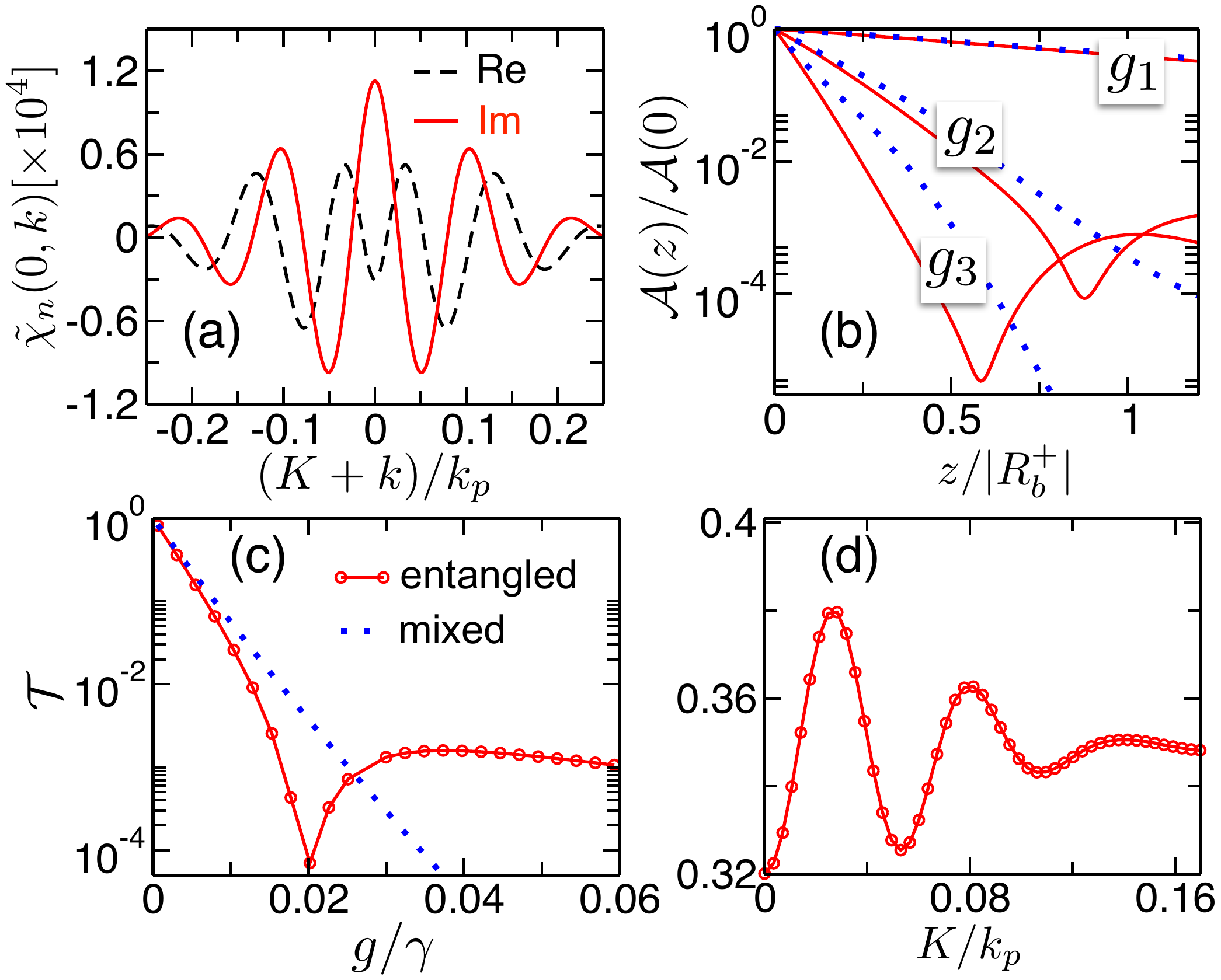}
\caption{(a) Real (dashed) and imaginary (solid) part of the non-local susceptibility $\tilde{\chi}_n(0,k)$ for $\Omega=1.6\gamma$ and $L=40\ \mu $m. (b) Probe light field density inside the medium (logarithmic scale) for different coupling constants: $g_1=4.0\times 10^{-3}\gamma$,  $g_2=4.0\times 10^{-2}\gamma$, and $g_3=1.0\times 10^{-1}\gamma$. The corresponding atom densities are $2.2\times10^{11}\ \text{cm}^{-3}$, $2.2\times 10^{12}\ \text{cm}^{-3}$ and $5.5\times 10^{12}\ \text{cm}^{-3}$, respectively. For each value of $g$ we plot $\mathcal{A}(z)/\mathcal{A}(0)$ for two initial situations: when the medium is entangled (solid curve) and when it is in a mixed state (dotted curve). (c) Photon transmission (logarithmic scale) as a function of $g$ in the phase-matched situation ($K=0$). (d) Transmission as a function of $K$. The corresponding coupling strength is $g_1$. Other parameters used for producing data in (b)-(d) are $\Omega=0.3\gamma$ and $L=17.6\,\mu $m. The corresponding blockade radii are $|R_{b}^{-}|=10.4\ \mu$m and $|R_{b}^{+}|=14.7\ \mu$m.}
\label{fig:transmission}
\end{figure}

\indent{\it Light propagation in a one dimensional gas.---} We study now a one-dimensional ensemble of atoms of length $L$, as shown in Fig.~\ref{fig:system}b, and a spatially homogeneous spinwave, i.e., with coefficients $A_j=1/\sqrt{N}$. Employing a continuum description we obtain the following polarization of this medium:
\begin{eqnarray}
\label{eq:gascoherence}
\mathcal{P}(z,t)&=&\epsilon_0\left[\chi_l(z)\mathcal{E}(z,t)+\int_L\chi_n(z,z')\mathcal{E}(z',t)dz'\right],
\end{eqnarray}
where $\epsilon_0$ is the vacuum permeability, and $\chi_l(z)=g\int_{L}\alpha^+_{zz'}\rho_s(z')dz'$ and
$\chi_n(z,z')=g\alpha^-_{zz'}e^{i\Phi_{zz'}}\rho_s(z')$
are the local and non-local susceptibilities, respectively. They depend both on the coupling constant $g=|\mu_{12}|^2N\rho_s(z)/\epsilon_0$ and the atom density $\rho_s(z)$ that is normalized according to $\int_L\rho_s(z)dz=1$.
Note, that all vectorial quantities have been replaced by their projection on the longitudinal axis $z$.

The propagation of the probe field through the medium is governed by the wave equation
$\left(\frac{\partial}{\partial z}+\frac{1}{c}\frac{\partial}{\partial
t}\right)\mathcal{E}(z,t)=\frac{ik_p}{2\epsilon_0}\mathcal{P}(z,t)$,
where $c$ is the light speed in vacuum ~\cite{gorshkov_photon-photon_2011,peyronel_2012}. The basic physics of the propagation can be understood by neglecting boundary effects and by performing a Fourier transform of Eq.~(\ref{eq:gascoherence}). This yields the susceptibility in $k$-space which consists of four terms: $\tilde{\chi}(\delta,k)=\tilde{\chi}_l^+(\delta)+\tilde{\chi}_l^-(\delta)+\tilde{\chi}_{n}^+(\delta,k)-\tilde{\chi}_{n}^-(\delta,k)$. Here
\begin{eqnarray}
\label{eq:susceptibility}
 \tilde{\chi}_l^{\pm}(\delta)&=&g\frac{\pi R_{b}^{\pm}\rho_{s}\Omega^{2}-3i\delta\Gamma}{3i\Gamma(2i\delta\Gamma-\Omega^2)},\\
 \tilde{\chi}_n^{\pm}(\delta,k)&=&g\frac{\sqrt{2\pi}}{12i\Gamma}\frac{ R_{b}^{\pm}\rho_{s}\Omega^{2}}{2i\delta\Gamma-\Omega^2}\left(\zeta_\pm+A\zeta_\pm^A+A^{*}\zeta_\pm^{A^{*}}\right),
\end{eqnarray}
which depends on $\zeta_\pm=e^{-R_b^{\pm}|k+K|}$, the complex number $A=(1+i\sqrt{3})/2$, and the two blockade radii \cite{gorshkov_photon-photon_2011} $|R_b^{\pm}|$, where
$R_b^{\pm}=[2\Gamma(C_d \pm C_e)/(\Omega^2-2i\delta\Gamma)]^{1/6}$.

This result exhibits three novel features due to the non-local component $\tilde{\chi}_n(\delta,k)\equiv\tilde{\chi}^+_n(\delta,k)-\tilde{\chi}_n^-(\delta,k)$. First, there is an explicit dependence on the wavenumber $k$ which is
shown in Fig.~\ref{fig:transmission}a. Second, the imaginary part of the non-local susceptibility becomes negative at certain values of $k$ which indicates a gain of probe field strength. This is a direct consequence of the fact that the polarization in one part of the medium acts as a source for other parts of the medium, analogous to the previously discussed case of two atoms. Note, however, that the imaginary part of the combined local and non-local susceptibilities is always greater than or equal to zero such that there is no net gain. Third, $\tilde{\chi}_n(\delta,k)$ decays exponentially with the wavevector $|k+K|$. Hence it drops rapidly when $|k+K|\gg 1/|R_{b}^{\pm}|$.

In the following we perform a numerical study in which we consider a resonant probe field pulse ($\delta=0$) and a gas of rubidium atoms with the two relevant Rydberg states $|83S,J=m_J=1/2\rangle\equiv |4\rangle$ and $|84S,J=m_J=1/2\rangle\equiv |3\rangle$. The corresponding dispersion coefficients are $C_d=9.7\times 10^{3}$ GHz$\,\mu$m$^6$ and $C_e=7.5\times10^{3}$ GHz$\,\mu$m$^6$. We first consider the phase-matched case (vanishing relative wavevector, $K=0$) for which we plot in Fig.~\ref{fig:transmission}b the time-integrated
density of the probe field $\mathcal{A}(z)=\int |\mathcal{E}(z,t)|^2dt$ for different values of the coupling constants $g$. For comparison we also provide data for an initial state which is an incoherent mixture of all configurations that contain a single atom in state $|3\rangle$ and all others in $|1\rangle$. This ``standard'' scenario has been much discussed in the recent literature, see e.g. Refs.~\cite{gorshkov_photon-photon_2011,olmos_2011,gunter_2012,gunter_observing_2013}. In Fig.~\ref{fig:transmission}b we find that for weak couplings, $\mathcal{A}(z)$ decays exponentially inside the medium no matter whether the initial state is mixed or a coherent superposition. With increasing coupling $g$, however, $\mathcal{A}(z)$ displays a non-exponential and non-monotonic decay in the case of an entangled initial state. The increase of $\mathcal{A}(z)$ at larger distances is a direct consequence of the negative imaginary part of the non-local susceptibility and the resulting gain. A similar effect is observed in the transmission, i.e., the mean number of photons transmitted through the medium (one photon at a time), defined by $\mathcal{T}=\mathcal{A}(L)/\mathcal{A}(0)$. The corresponding data as a function of $g$, shown in Fig.~\ref{fig:transmission}c, again displays a non-monotonic behaviour and is increasing beyond $g=0.019$. No such feature is present for a mixed initial state~\cite{gorshkov_photon-photon_2011}. The momentum dependence of the susceptibility is also evidenced in Fig.~\ref{fig:transmission}d which shows the transmission as a function of the relative wavevector $K$. The observed damped oscillations with increasing $K$ are a consequence of the oscillating behaviour of the non-local susceptibility that was previously discussed in Fig.~\ref{fig:transmission}a.
\begin{figure}
\centering
\includegraphics*[width=0.92\columnwidth]{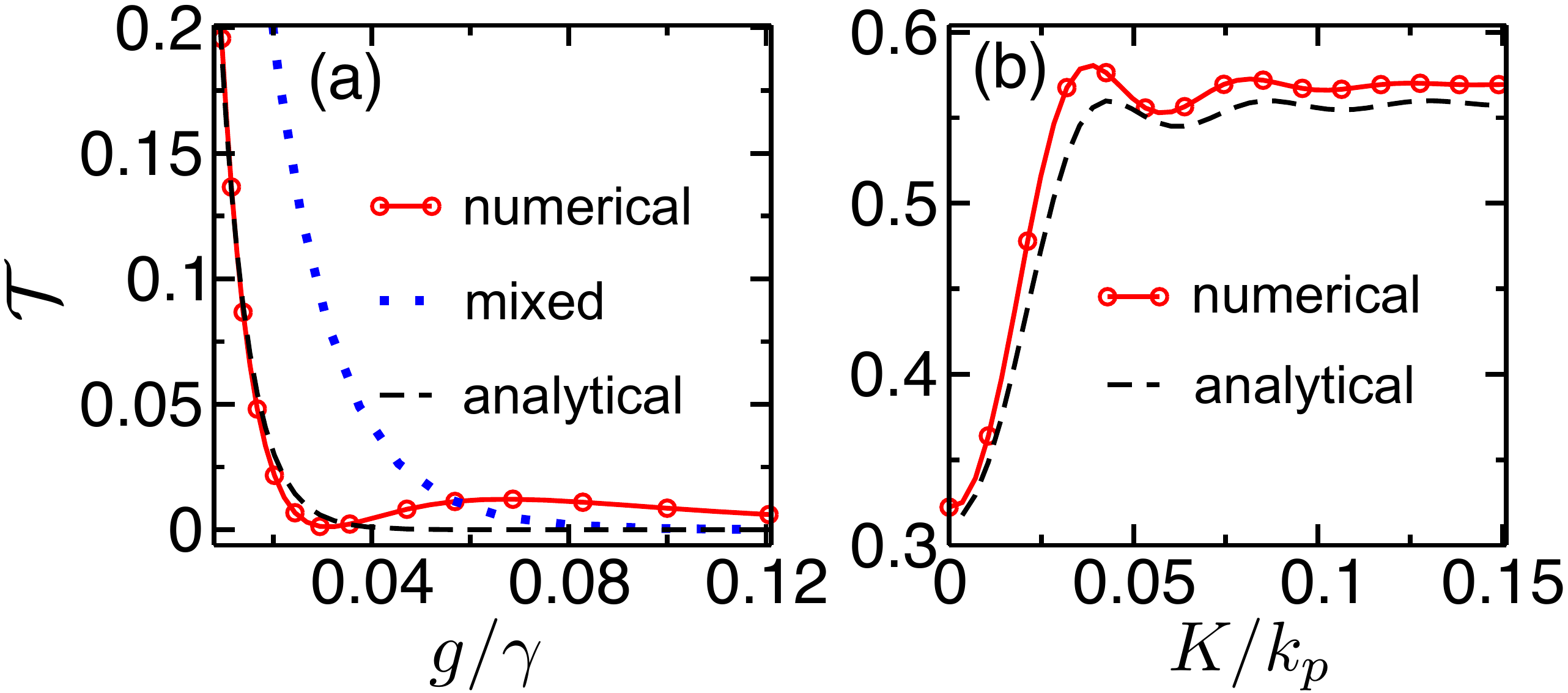}
\caption{Transmission $\mathcal{T}$ as a function of the coupling strength $g$ [panel (a)] and the relative wavevector $K$ [panel (b)] for the case of equal van-der-Waals and exchange interaction coefficients $C_d=C_e=9.7\times 10^3$ GHz$\,\mu\text{m}^6$. The other parameters are the same as in Fig.~\ref{fig:transmission}d. As shown in (a), the absorption of the entangled medium is greatly enhanced with respect to the mixed case. The analytical approximation of the transmission $\mathcal{T}_\mathrm{a}$ performs well for sufficiently weak coupling.}
\label{fig:spinwave}
\end{figure}

Finally, let us remark that an interesting scenario emerges when the dispersion coefficients of the van-der-Waals and the exchange interaction are equal, i.e., $C_d=C_e$, which can potentially be approximately achieved by tailoring the interaction potential through external fields~\cite{anderson_resonant_1998,vogt_2006,ryabtsev_2010,nipper_highly_2012}. As shown in Fig.~\ref{fig:spinwave}a the gain effect is here about ten times stronger as compared to the case in Fig.~\ref{fig:transmission}c. The reason for this enhancement is that for $C_d=C_e$ all anti-symmetric excitonic states, $\left(|1\cdots3_j\cdots 4_k\cdots\rangle-|1\cdots4_j\cdots
3_k\cdots\rangle\right)/\sqrt{2}$, have zero interaction energy. Coupling to these states therefore reduces absorption (c.f. the two-atom case in Fig.~\ref{fig:susceptibility}).

For $C_d=C_e$ an approximate analytical solution for the probe field amplitude can be derived. It does not capture the gain but demonstrates the enhanced absorption with respect to conventional Rydberg EIT starting from a mixed state: assuming that the medium is fully blockaded ($L\lesssim |R_b^+|$) such that the polarization is space-independent~\cite{gorshkov_photon-photon_2011}, i.e., $\alpha_{zz'}^{\pm}\approx i/2\gamma$ and applying the local field approximation~\cite{sevincli_nonlocal_2011}, the electric field in the medium (in the space domain) is
\begin{eqnarray}
\label{eq:analytical}
\mathcal{E}_a(z)=\mathcal{E}(0)e^{-\frac{k_p\,g}{4\gamma}\left(1+\frac{4\sin\frac{KL}{2}}{K^2Lz}\sin\frac{Kz}{2}e^{iK\frac{z-L}{2}}\right)z},
\end{eqnarray}
where $\mathcal{E}(0)$ is the electric field at $z=0$. From this expression we find that the transmission is given by
$\mathcal{T}_\mathrm{a}=e^{-k_pgL[1+(2-2\cos KL)/(KL)^2]/(2\gamma)}$, which indeed agrees well with the numerical solution shown in Fig.~\ref{fig:spinwave}b until the coupling exceeds $g_0=2\ln(2)\gamma/k_pL$. Note that a mixed initial state leads under the same approximations to a transmission that is $\mathcal{T}^\mathrm{mixed}_\mathrm{a}=e^{-k_pgL/(2\gamma)}$ and hence substantially weaker absorption. The analytical form of $\mathcal{T}_a$ once again highlights the non-trivial dependence of the transmission on the relative wavevector $K$ (Fig.~\ref{fig:spinwave}b) and shows that $\mathcal{T}_a$ saturates at $\mathcal{T}^\mathrm{mixed}_\mathrm{a}$ for large $K$.

\indent{\it Summary and outlook.---} We have studied the dynamics of Rydberg EIT in a setting with two relevant excited states \cite{Ditzhuijzen08,gunter_observing_2013} where exchange leads to novel features in the atom-light interaction. If the initial state is a spinwave, our results predict non-local light propagation and a greatly enhanced absorption. These effects can be of relevance for applications such as optical photon switches \cite{baur_single_2014}, routers \cite{RoccaPRL2009}, and gates \cite{gorshkov_photon-photon_2011}. Furthermore, excitonic dynamics in multi-component Rydberg gases \cite{Ditzhuijzen08} provide a powerful
new system to study energy transport phenomena in complex networks
\cite{gunter_observing_2013}. Our results, in particular the
$k$-dependence of the transmission, open novel approaches for probing
this rich dynamics.

\begin{acknowledgements}
\textit{Acknowledgements.---} We acknowledge helpful discussions with S. Weber, H. Gorniaczyk, C. Tresp, S, Genway and M. Hush. The research leading to these results has received funding from the European Research Council under the European Union's Seventh Framework Programme (FP/2007-2013) / ERC Grant Agreement n. 335266 (ESCQUMA), the EU-FET grant QuILMI 295293, the ERA-NET CHIST-ERA (R-ION consortium) and from the University of Nottingham. S.H. is funded by the German Research Foundation through project HO 4787/1-1.
\end{acknowledgements}


\begin{thebibliography}{39}%
\makeatletter
\providecommand \@ifxundefined [1]{%
 \@ifx{#1\undefined}
}%
\providecommand \@ifnum [1]{%
 \ifnum #1\expandafter \@firstoftwo
 \else \expandafter \@secondoftwo
 \fi
}%
\providecommand \@ifx [1]{%
 \ifx #1\expandafter \@firstoftwo
 \else \expandafter \@secondoftwo
 \fi
}%
\providecommand \natexlab [1]{#1}%
\providecommand \enquote  [1]{``#1''}%
\providecommand \bibnamefont  [1]{#1}%
\providecommand \bibfnamefont [1]{#1}%
\providecommand \citenamefont [1]{#1}%
\providecommand \href@noop [0]{\@secondoftwo}%
\providecommand \href [0]{\begingroup \@sanitize@url \@href}%
\providecommand \@href[1]{\@@startlink{#1}\@@href}%
\providecommand \@@href[1]{\endgroup#1\@@endlink}%
\providecommand \@sanitize@url [0]{\catcode `\\12\catcode `\$12\catcode
  `\&12\catcode `\#12\catcode `\^12\catcode `\_12\catcode `\%12\relax}%
\providecommand \@@startlink[1]{}%
\providecommand \@@endlink[0]{}%
\providecommand \url  [0]{\begingroup\@sanitize@url \@url }%
\providecommand \@url [1]{\endgroup\@href {#1}{\urlprefix }}%
\providecommand \urlprefix  [0]{URL }%
\providecommand \Eprint [0]{\href }%
\providecommand \doibase [0]{http://dx.doi.org/}%
\providecommand \selectlanguage [0]{\@gobble}%
\providecommand \bibinfo  [0]{\@secondoftwo}%
\providecommand \bibfield  [0]{\@secondoftwo}%
\providecommand \translation [1]{[#1]}%
\providecommand \BibitemOpen [0]{}%
\providecommand \bibitemStop [0]{}%
\providecommand \bibitemNoStop [0]{.\EOS\space}%
\providecommand \EOS [0]{\spacefactor3000\relax}%
\providecommand \BibitemShut  [1]{\csname bibitem#1\endcsname}%
\let\auto@bib@innerbib\@empty
\bibitem [{\citenamefont {Fleischhauer}\ \emph {et~al.}(2005)\citenamefont
  {Fleischhauer}, \citenamefont {Imamoglu},\ and\ \citenamefont
  {Marangos}}]{fleischhauer_EIT_2005}%
  \BibitemOpen
  \bibfield  {author} {\bibinfo {author} {\bibfnamefont {M.}~\bibnamefont
  {Fleischhauer}}, \bibinfo {author} {\bibfnamefont {A.}~\bibnamefont
  {Imamoglu}}, \ and\ \bibinfo {author} {\bibfnamefont {J.~P.}\ \bibnamefont
  {Marangos}},\ }\href {\doibase 10.1103/RevModPhys.77.633} {\bibfield
  {journal} {\bibinfo  {journal} {Rev. Mod. Phys.}\ }\textbf {\bibinfo {volume}
  {77}},\ \bibinfo {pages} {633} (\bibinfo {year} {2005})}\BibitemShut
  {NoStop}%
\bibitem [{\citenamefont {Saffman}\ \emph {et~al.}(2010)\citenamefont
  {Saffman}, \citenamefont {Walker},\ and\ \citenamefont
  {M{\o}lmer}}]{saffman_quantum_2010}%
  \BibitemOpen
  \bibfield  {author} {\bibinfo {author} {\bibfnamefont {M.}~\bibnamefont
  {Saffman}}, \bibinfo {author} {\bibfnamefont {T.~G.}\ \bibnamefont {Walker}},
  \ and\ \bibinfo {author} {\bibfnamefont {K.}~\bibnamefont {M{\o}lmer}},\
  }\href {\doibase 10.1103/RevModPhys.82.2313} {\bibfield  {journal} {\bibinfo
  {journal} {Rev. Mod. Phys.}\ }\textbf {\bibinfo {volume} {82}},\ \bibinfo
  {pages} {2313} (\bibinfo {year} {2010})}\BibitemShut {NoStop}%
\bibitem [{\citenamefont {Pritchard}\ \emph {et~al.}(2010)\citenamefont
  {Pritchard}, \citenamefont {Maxwell}, \citenamefont {Gauguet}, \citenamefont
  {Weatherill}, \citenamefont {Jones},\ and\ \citenamefont
  {Adams}}]{pritchard_cooperative_2010}%
  \BibitemOpen
  \bibfield  {author} {\bibinfo {author} {\bibfnamefont {J.~D.}\ \bibnamefont
  {Pritchard}}, \bibinfo {author} {\bibfnamefont {D.}~\bibnamefont {Maxwell}},
  \bibinfo {author} {\bibfnamefont {A.}~\bibnamefont {Gauguet}}, \bibinfo
  {author} {\bibfnamefont {K.~J.}\ \bibnamefont {Weatherill}}, \bibinfo
  {author} {\bibfnamefont {M.~P.~A.}\ \bibnamefont {Jones}}, \ and\ \bibinfo
  {author} {\bibfnamefont {C.~S.}\ \bibnamefont {Adams}},\ }\href {\doibase
  10.1103/PhysRevLett.105.193603} {\bibfield  {journal} {\bibinfo  {journal}
  {Phys. Rev. Lett.}\ }\textbf {\bibinfo {volume} {105}},\ \bibinfo {pages}
  {193603} (\bibinfo {year} {2010})}\BibitemShut {NoStop}%
\bibitem [{\citenamefont {Friedler}\ \emph {et~al.}(2005)\citenamefont
  {Friedler}, \citenamefont {Petrosyan}, \citenamefont {Fleischhauer},\ and\
  \citenamefont {Kurizki}}]{friedler_long-range_2005}%
  \BibitemOpen
  \bibfield  {author} {\bibinfo {author} {\bibfnamefont {I.}~\bibnamefont
  {Friedler}}, \bibinfo {author} {\bibfnamefont {D.}~\bibnamefont {Petrosyan}},
  \bibinfo {author} {\bibfnamefont {M.}~\bibnamefont {Fleischhauer}}, \ and\
  \bibinfo {author} {\bibfnamefont {G.}~\bibnamefont {Kurizki}},\ }\href
  {\doibase 10.1103/PhysRevA.72.043803} {\bibfield  {journal} {\bibinfo
  {journal} {Phys. Rev. A}\ }\textbf {\bibinfo {volume} {72}},\ \bibinfo
  {pages} {043803} (\bibinfo {year} {2005})}\BibitemShut {NoStop}%
\bibitem [{\citenamefont {Ates}\ \emph {et~al.}(2011)\citenamefont {Ates},
  \citenamefont {Sevin\ifmmode~\mbox{\c{c}}\else \c{c}\fi{}li},\ and\
  \citenamefont {Pohl}}]{ates_eit_2011}%
  \BibitemOpen
  \bibfield  {author} {\bibinfo {author} {\bibfnamefont {C.}~\bibnamefont
  {Ates}}, \bibinfo {author} {\bibfnamefont {S.}~\bibnamefont
  {Sevin\ifmmode~\mbox{\c{c}}\else \c{c}\fi{}li}}, \ and\ \bibinfo {author}
  {\bibfnamefont {T.}~\bibnamefont {Pohl}},\ }\href {\doibase
  10.1103/PhysRevA.83.041802} {\bibfield  {journal} {\bibinfo  {journal} {Phys.
  Rev. A}\ }\textbf {\bibinfo {volume} {83}},\ \bibinfo {pages} {041802}
  (\bibinfo {year} {2011})}\BibitemShut {NoStop}%
\bibitem [{\citenamefont {Olmos}\ \emph {et~al.}(2011)\citenamefont {Olmos},
  \citenamefont {Li}, \citenamefont {Hofferberth},\ and\ \citenamefont
  {Lesanovsky}}]{olmos_2011}%
  \BibitemOpen
  \bibfield  {author} {\bibinfo {author} {\bibfnamefont {B.}~\bibnamefont
  {Olmos}}, \bibinfo {author} {\bibfnamefont {W.}~\bibnamefont {Li}}, \bibinfo
  {author} {\bibfnamefont {S.}~\bibnamefont {Hofferberth}}, \ and\ \bibinfo
  {author} {\bibfnamefont {I.}~\bibnamefont {Lesanovsky}},\ }\href {\doibase
  10.1103/PhysRevA.84.041607} {\bibfield  {journal} {\bibinfo  {journal} {Phys.
  Rev. A}\ }\textbf {\bibinfo {volume} {84}},\ \bibinfo {pages} {041607}
  (\bibinfo {year} {2011})}\BibitemShut {NoStop}%
\bibitem [{\citenamefont {G\"unter}\ \emph {et~al.}(2012)\citenamefont
  {G\"unter}, \citenamefont {Robert-de Saint-Vincent}, \citenamefont {Schempp},
  \citenamefont {Hofmann}, \citenamefont {Whitlock},\ and\ \citenamefont
  {Weidem\"uller}}]{gunter_2012}%
  \BibitemOpen
  \bibfield  {author} {\bibinfo {author} {\bibfnamefont {G.}~\bibnamefont
  {G\"unter}}, \bibinfo {author} {\bibfnamefont {M.}~\bibnamefont {Robert-de
  Saint-Vincent}}, \bibinfo {author} {\bibfnamefont {H.}~\bibnamefont
  {Schempp}}, \bibinfo {author} {\bibfnamefont {C.~S.}\ \bibnamefont
  {Hofmann}}, \bibinfo {author} {\bibfnamefont {S.}~\bibnamefont {Whitlock}}, \
  and\ \bibinfo {author} {\bibfnamefont {M.}~\bibnamefont {Weidem\"uller}},\
  }\href {\doibase 10.1103/PhysRevLett.108.013002} {\bibfield  {journal}
  {\bibinfo  {journal} {Phys. Rev. Lett.}\ }\textbf {\bibinfo {volume} {108}},\
  \bibinfo {pages} {013002} (\bibinfo {year} {2012})}\BibitemShut {NoStop}%
\bibitem [{\citenamefont {Dudin}\ and\ \citenamefont
  {Kuzmich}(2012)}]{dudin_strongly_2012}%
  \BibitemOpen
  \bibfield  {author} {\bibinfo {author} {\bibfnamefont {Y.~O.}\ \bibnamefont
  {Dudin}}\ and\ \bibinfo {author} {\bibfnamefont {A.}~\bibnamefont
  {Kuzmich}},\ }\href {\doibase 10.1126/science.1217901} {\bibfield  {journal}
  {\bibinfo  {journal} {Science}\ }\textbf {\bibinfo {volume} {336}},\ \bibinfo
  {pages} {887–} (\bibinfo {year} {2012})}\BibitemShut {NoStop}%
\bibitem [{\citenamefont {Yan}\ \emph {et~al.}(2012)\citenamefont {Yan},
  \citenamefont {Liu}, \citenamefont {Bao}, \citenamefont {Fu},\ and\
  \citenamefont {Wu}}]{yan_2012}%
  \BibitemOpen
  \bibfield  {author} {\bibinfo {author} {\bibfnamefont {D.}~\bibnamefont
  {Yan}}, \bibinfo {author} {\bibfnamefont {Y.-M.}\ \bibnamefont {Liu}},
  \bibinfo {author} {\bibfnamefont {Q.-Q.}\ \bibnamefont {Bao}}, \bibinfo
  {author} {\bibfnamefont {C.-B.}\ \bibnamefont {Fu}}, \ and\ \bibinfo {author}
  {\bibfnamefont {J.-H.}\ \bibnamefont {Wu}},\ }\href {\doibase
  10.1103/PhysRevA.86.023828} {\bibfield  {journal} {\bibinfo  {journal} {Phys.
  Rev. A}\ }\textbf {\bibinfo {volume} {86}},\ \bibinfo {pages} {023828}
  (\bibinfo {year} {2012})}\BibitemShut {NoStop}%
\bibitem [{\citenamefont {G\"arttner}\ and\ \citenamefont
  {Evers}(2013)}]{garttner_2013}%
  \BibitemOpen
  \bibfield  {author} {\bibinfo {author} {\bibfnamefont {M.}~\bibnamefont
  {G\"arttner}}\ and\ \bibinfo {author} {\bibfnamefont {J.}~\bibnamefont
  {Evers}},\ }\href {\doibase 10.1103/PhysRevA.88.033417} {\bibfield  {journal}
  {\bibinfo  {journal} {Phys. Rev. A}\ }\textbf {\bibinfo {volume} {88}},\
  \bibinfo {pages} {033417} (\bibinfo {year} {2013})}\BibitemShut {NoStop}%
\bibitem [{\citenamefont {Gorshkov}\ \emph {et~al.}(2011)\citenamefont
  {Gorshkov}, \citenamefont {Otterbach}, \citenamefont {Fleischhauer},
  \citenamefont {Pohl},\ and\ \citenamefont
  {Lukin}}]{gorshkov_photon-photon_2011}%
  \BibitemOpen
  \bibfield  {author} {\bibinfo {author} {\bibfnamefont {A.~V.}\ \bibnamefont
  {Gorshkov}}, \bibinfo {author} {\bibfnamefont {J.}~\bibnamefont {Otterbach}},
  \bibinfo {author} {\bibfnamefont {M.}~\bibnamefont {Fleischhauer}}, \bibinfo
  {author} {\bibfnamefont {T.}~\bibnamefont {Pohl}}, \ and\ \bibinfo {author}
  {\bibfnamefont {M.~D.}\ \bibnamefont {Lukin}},\ }\href {\doibase
  10.1103/PhysRevLett.107.133602} {\bibfield  {journal} {\bibinfo  {journal}
  {Phys. Rev. Lett.}\ }\textbf {\bibinfo {volume} {107}},\ \bibinfo {pages}
  {133602} (\bibinfo {year} {2011})}\BibitemShut {NoStop}%
\bibitem [{\citenamefont {Peyronel}\ \emph {et~al.}(2012)\citenamefont
  {Peyronel}, \citenamefont {Firstenberg}, \citenamefont {Liang}, \citenamefont
  {Hofferberth}, \citenamefont {Gorshkov}, \citenamefont {Pohl}, \citenamefont
  {Lukin},\ and\ \citenamefont {Vuletic}}]{peyronel_2012}%
  \BibitemOpen
  \bibfield  {author} {\bibinfo {author} {\bibfnamefont {T.}~\bibnamefont
  {Peyronel}}, \bibinfo {author} {\bibfnamefont {O.}~\bibnamefont
  {Firstenberg}}, \bibinfo {author} {\bibfnamefont {Q.-Y.}\ \bibnamefont
  {Liang}}, \bibinfo {author} {\bibfnamefont {S.}~\bibnamefont {Hofferberth}},
  \bibinfo {author} {\bibfnamefont {A.~V.}\ \bibnamefont {Gorshkov}}, \bibinfo
  {author} {\bibfnamefont {T.}~\bibnamefont {Pohl}}, \bibinfo {author}
  {\bibfnamefont {M.~D.}\ \bibnamefont {Lukin}}, \ and\ \bibinfo {author}
  {\bibfnamefont {V.}~\bibnamefont {Vuletic}},\ }\href
  {http://dx.doi.org/10.1038/nature11361} {\bibfield  {journal} {\bibinfo
  {journal} {Nature}\ }\textbf {\bibinfo {volume} {488}},\ \bibinfo {pages}
  {57} (\bibinfo {year} {2012})}\BibitemShut {NoStop}%
\bibitem [{\citenamefont {Gorshkov}\ \emph {et~al.}(2013)\citenamefont
  {Gorshkov}, \citenamefont {Nath},\ and\ \citenamefont
  {Pohl}}]{gorshkov_dissipative_2013}%
  \BibitemOpen
  \bibfield  {author} {\bibinfo {author} {\bibfnamefont {A.~V.}\ \bibnamefont
  {Gorshkov}}, \bibinfo {author} {\bibfnamefont {R.}~\bibnamefont {Nath}}, \
  and\ \bibinfo {author} {\bibfnamefont {T.}~\bibnamefont {Pohl}},\ }\href
  {\doibase 10.1103/PhysRevLett.110.153601} {\bibfield  {journal} {\bibinfo
  {journal} {Phys. Rev. Lett.}\ }\textbf {\bibinfo {volume} {110}},\ \bibinfo
  {pages} {153601} (\bibinfo {year} {2013})}\BibitemShut {NoStop}%
\bibitem [{\citenamefont {Baur}\ \emph {et~al.}(2014)\citenamefont {Baur},
  \citenamefont {Tiarks}, \citenamefont {Rempe},\ and\ \citenamefont
  {D\"urr}}]{baur_single_2014}%
  \BibitemOpen
  \bibfield  {author} {\bibinfo {author} {\bibfnamefont {S.}~\bibnamefont
  {Baur}}, \bibinfo {author} {\bibfnamefont {D.}~\bibnamefont {Tiarks}},
  \bibinfo {author} {\bibfnamefont {G.}~\bibnamefont {Rempe}}, \ and\ \bibinfo
  {author} {\bibfnamefont {S.}~\bibnamefont {D\"urr}},\ }\href {\doibase
  10.1103/PhysRevLett.112.073901} {\bibfield  {journal} {\bibinfo  {journal}
  {Phys. Rev. Lett.}\ }\textbf {\bibinfo {volume} {112}},\ \bibinfo {pages}
  {073901} (\bibinfo {year} {2014})}\BibitemShut {NoStop}%
\bibitem [{\citenamefont {Firstenberg}\ \emph {et~al.}(2013)\citenamefont
  {Firstenberg}, \citenamefont {Peyronel}, \citenamefont {Liang}, \citenamefont
  {Gorshkov}, \citenamefont {Lukin},\ and\ \citenamefont
  {Vuleti{\'c}}}]{firstenberg_attractive_2013}%
  \BibitemOpen
  \bibfield  {author} {\bibinfo {author} {\bibfnamefont {O.}~\bibnamefont
  {Firstenberg}}, \bibinfo {author} {\bibfnamefont {T.}~\bibnamefont
  {Peyronel}}, \bibinfo {author} {\bibfnamefont {Q.-Y.}\ \bibnamefont {Liang}},
  \bibinfo {author} {\bibfnamefont {A.~V.}\ \bibnamefont {Gorshkov}}, \bibinfo
  {author} {\bibfnamefont {M.~D.}\ \bibnamefont {Lukin}}, \ and\ \bibinfo
  {author} {\bibfnamefont {V.}~\bibnamefont {Vuleti{\'c}}},\ }\href {\doibase
  10.1038/nature12512} {\bibfield  {journal} {\bibinfo  {journal} {Nature}\
  }\textbf {\bibinfo {volume} {502}},\ \bibinfo {pages} {71} (\bibinfo {year}
  {2013})}\BibitemShut {NoStop}%
\bibitem [{\citenamefont {Bienias}\ \emph {et~al.}(2014)\citenamefont
  {Bienias}, \citenamefont {Choi}, \citenamefont {Firstenberg}, \citenamefont
  {Maghrebi}, \citenamefont {Gullans}, \citenamefont {Lukin}, \citenamefont
  {Gorshkov},\ and\ \citenamefont {B\"uchler}}]{Bienias14}%
  \BibitemOpen
  \bibfield  {author} {\bibinfo {author} {\bibfnamefont {P.}~\bibnamefont
  {Bienias}}, \bibinfo {author} {\bibfnamefont {S.}~\bibnamefont {Choi}},
  \bibinfo {author} {\bibfnamefont {O.}~\bibnamefont {Firstenberg}}, \bibinfo
  {author} {\bibfnamefont {M.~F.}\ \bibnamefont {Maghrebi}}, \bibinfo {author}
  {\bibfnamefont {M.}~\bibnamefont {Gullans}}, \bibinfo {author} {\bibfnamefont
  {M.~D.}\ \bibnamefont {Lukin}}, \bibinfo {author} {\bibfnamefont
  {A.}~\bibnamefont {Gorshkov}}, \ and\ \bibinfo {author} {\bibfnamefont
  {H.~P.}\ \bibnamefont {B\"uchler}},\ }\href@noop {} {\bibfield  {journal}
  {\bibinfo  {journal} {{arXiv:1402.7333v1}}\ } (\bibinfo {year}
  {2014})}\BibitemShut {NoStop}%
\bibitem [{\citenamefont {Otterbach}\ \emph {et~al.}(2013)\citenamefont
  {Otterbach}, \citenamefont {Moos}, \citenamefont {Muth},\ and\ \citenamefont
  {Fleischhauer}}]{otterbach_wigner_2013}%
  \BibitemOpen
  \bibfield  {author} {\bibinfo {author} {\bibfnamefont {J.}~\bibnamefont
  {Otterbach}}, \bibinfo {author} {\bibfnamefont {M.}~\bibnamefont {Moos}},
  \bibinfo {author} {\bibfnamefont {D.}~\bibnamefont {Muth}}, \ and\ \bibinfo
  {author} {\bibfnamefont {M.}~\bibnamefont {Fleischhauer}},\ }\href {\doibase
  10.1103/PhysRevLett.111.113001} {\bibfield  {journal} {\bibinfo  {journal}
  {Phys. Rev. Lett.}\ }\textbf {\bibinfo {volume} {111}},\ \bibinfo {pages}
  {113001} (\bibinfo {year} {2013})}\BibitemShut {NoStop}%
\bibitem [{\citenamefont {Sevin\ifmmode~\mbox{\c{c}}\else \c{c}\fi{}li}\ \emph
  {et~al.}(2011)\citenamefont {Sevin\ifmmode~\mbox{\c{c}}\else \c{c}\fi{}li},
  \citenamefont {Henkel}, \citenamefont {Ates},\ and\ \citenamefont
  {Pohl}}]{sevincli_nonlocal_2011}%
  \BibitemOpen
  \bibfield  {author} {\bibinfo {author} {\bibfnamefont {S.}~\bibnamefont
  {Sevin\ifmmode~\mbox{\c{c}}\else \c{c}\fi{}li}}, \bibinfo {author}
  {\bibfnamefont {N.}~\bibnamefont {Henkel}}, \bibinfo {author} {\bibfnamefont
  {C.}~\bibnamefont {Ates}}, \ and\ \bibinfo {author} {\bibfnamefont
  {T.}~\bibnamefont {Pohl}},\ }\href {\doibase 10.1103/PhysRevLett.107.153001}
  {\bibfield  {journal} {\bibinfo  {journal} {Phys. Rev. Lett.}\ }\textbf
  {\bibinfo {volume} {107}},\ \bibinfo {pages} {153001} (\bibinfo {year}
  {2011})}\BibitemShut {NoStop}%
\bibitem [{\citenamefont {Petrosyan}\ \emph {et~al.}(2011)\citenamefont
  {Petrosyan}, \citenamefont {Otterbach},\ and\ \citenamefont
  {Fleischhauer}}]{petrosyan_electromagnetically_2011}%
  \BibitemOpen
  \bibfield  {author} {\bibinfo {author} {\bibfnamefont {D.}~\bibnamefont
  {Petrosyan}}, \bibinfo {author} {\bibfnamefont {J.}~\bibnamefont
  {Otterbach}}, \ and\ \bibinfo {author} {\bibfnamefont {M.}~\bibnamefont
  {Fleischhauer}},\ }\href {\doibase 10.1103/PhysRevLett.107.213601} {\bibfield
   {journal} {\bibinfo  {journal} {Phys. Rev. Lett.}\ }\textbf {\bibinfo
  {volume} {107}},\ \bibinfo {pages} {213601} (\bibinfo {year}
  {2011})}\BibitemShut {NoStop}%
\bibitem [{\citenamefont {Pohl}\ \emph {et~al.}(2010)\citenamefont {Pohl},
  \citenamefont {Demler},\ and\ \citenamefont {Lukin}}]{pohl_dynamical_2009}%
  \BibitemOpen
  \bibfield  {author} {\bibinfo {author} {\bibfnamefont {T.}~\bibnamefont
  {Pohl}}, \bibinfo {author} {\bibfnamefont {E.}~\bibnamefont {Demler}}, \ and\
  \bibinfo {author} {\bibfnamefont {M.~D.}\ \bibnamefont {Lukin}},\ }\href
  {\doibase 10.1103/PhysRevLett.104.043002} {\bibfield  {journal} {\bibinfo
  {journal} {Phys. Rev. Lett.}\ }\textbf {\bibinfo {volume} {104}},\ \bibinfo
  {pages} {043002} (\bibinfo {year} {2010})}\BibitemShut {NoStop}%
\bibitem [{\citenamefont {Lesanovsky}(2011)}]{lesanovsky_many-body_2011}%
  \BibitemOpen
  \bibfield  {author} {\bibinfo {author} {\bibfnamefont {I.}~\bibnamefont
  {Lesanovsky}},\ }\href {\doibase 10.1103/PhysRevLett.106.025301} {\bibfield
  {journal} {\bibinfo  {journal} {Phys. Rev. Lett.}\ }\textbf {\bibinfo
  {volume} {106}},\ \bibinfo {pages} {025301} (\bibinfo {year}
  {2011})}\BibitemShut {NoStop}%
\bibitem [{\citenamefont {Schau{\ss}}\ \emph {et~al.}(2012)\citenamefont
  {Schau{\ss}}, \citenamefont {Cheneau}, \citenamefont {Endres}, \citenamefont
  {Fukuhara}, \citenamefont {Hild}, \citenamefont {Omran}, \citenamefont
  {Pohl}, \citenamefont {Gross}, \citenamefont {Kuhr},\ and\ \citenamefont
  {Bloch}}]{schaus_observation_2012}%
  \BibitemOpen
  \bibfield  {author} {\bibinfo {author} {\bibfnamefont {P.}~\bibnamefont
  {Schau{\ss}}}, \bibinfo {author} {\bibfnamefont {M.}~\bibnamefont {Cheneau}},
  \bibinfo {author} {\bibfnamefont {M.}~\bibnamefont {Endres}}, \bibinfo
  {author} {\bibfnamefont {T.}~\bibnamefont {Fukuhara}}, \bibinfo {author}
  {\bibfnamefont {S.}~\bibnamefont {Hild}}, \bibinfo {author} {\bibfnamefont
  {A.}~\bibnamefont {Omran}}, \bibinfo {author} {\bibfnamefont
  {T.}~\bibnamefont {Pohl}}, \bibinfo {author} {\bibfnamefont {C.}~\bibnamefont
  {Gross}}, \bibinfo {author} {\bibfnamefont {S.}~\bibnamefont {Kuhr}}, \ and\
  \bibinfo {author} {\bibfnamefont {I.}~\bibnamefont {Bloch}},\ }\href
  {\doibase 10.1038/nature11596} {\bibfield  {journal} {\bibinfo  {journal}
  {Nature}\ }\textbf {\bibinfo {volume} {491}},\ \bibinfo {pages} {87}
  (\bibinfo {year} {2012})}\BibitemShut {NoStop}%
\bibitem [{\citenamefont {Anderson}\ \emph {et~al.}(1998)\citenamefont
  {Anderson}, \citenamefont {Veale},\ and\ \citenamefont
  {Gallagher}}]{anderson_resonant_1998}%
  \BibitemOpen
  \bibfield  {author} {\bibinfo {author} {\bibfnamefont {W.~R.}\ \bibnamefont
  {Anderson}}, \bibinfo {author} {\bibfnamefont {J.~R.}\ \bibnamefont {Veale}},
  \ and\ \bibinfo {author} {\bibfnamefont {T.~F.}\ \bibnamefont {Gallagher}},\
  }\href {\doibase 10.1103/PhysRevLett.80.249} {\bibfield  {journal} {\bibinfo
  {journal} {Phys. Rev. Lett.}\ }\textbf {\bibinfo {volume} {80}},\ \bibinfo
  {pages} {249} (\bibinfo {year} {1998})}\BibitemShut {NoStop}%
\bibitem [{\citenamefont {van Ditzhuijzen}\ \emph {et~al.}(2008)\citenamefont
  {van Ditzhuijzen}, \citenamefont {Koenderink}, \citenamefont {Hern\'andez},
  \citenamefont {Robicheaux}, \citenamefont {Noordam},\ and\ \citenamefont
  {van~den Heuvell}}]{Ditzhuijzen08}%
  \BibitemOpen
  \bibfield  {author} {\bibinfo {author} {\bibfnamefont {C.~S.~E.}\
  \bibnamefont {van Ditzhuijzen}}, \bibinfo {author} {\bibfnamefont {A.~F.}\
  \bibnamefont {Koenderink}}, \bibinfo {author} {\bibfnamefont {J.~V.}\
  \bibnamefont {Hern\'andez}}, \bibinfo {author} {\bibfnamefont
  {F.}~\bibnamefont {Robicheaux}}, \bibinfo {author} {\bibfnamefont {L.~D.}\
  \bibnamefont {Noordam}}, \ and\ \bibinfo {author} {\bibfnamefont {H.~B.
  v.~L.}\ \bibnamefont {van~den Heuvell}},\ }\href {\doibase
  10.1103/PhysRevLett.100.243201} {\bibfield  {journal} {\bibinfo  {journal}
  {Phys. Rev. Lett.}\ }\textbf {\bibinfo {volume} {100}},\ \bibinfo {pages}
  {243201} (\bibinfo {year} {2008})}\BibitemShut {NoStop}%
\bibitem [{\citenamefont {Nipper}\ \emph {et~al.}(2012)\citenamefont {Nipper},
  \citenamefont {Balewski}, \citenamefont {Krupp}, \citenamefont {Butscher},
  \citenamefont {L{\"o}w},\ and\ \citenamefont {Pfau}}]{nipper_highly_2012}%
  \BibitemOpen
  \bibfield  {author} {\bibinfo {author} {\bibfnamefont {J.}~\bibnamefont
  {Nipper}}, \bibinfo {author} {\bibfnamefont {J.~B.}\ \bibnamefont
  {Balewski}}, \bibinfo {author} {\bibfnamefont {A.~T.}\ \bibnamefont {Krupp}},
  \bibinfo {author} {\bibfnamefont {B.}~\bibnamefont {Butscher}}, \bibinfo
  {author} {\bibfnamefont {R.}~\bibnamefont {L{\"o}w}}, \ and\ \bibinfo
  {author} {\bibfnamefont {T.}~\bibnamefont {Pfau}},\ }\href {\doibase
  10.1103/PhysRevLett.108.113001} {\bibfield  {journal} {\bibinfo  {journal}
  {Phys. Rev. Lett.}\ }\textbf {\bibinfo {volume} {108}},\ \bibinfo {pages}
  {113001} (\bibinfo {year} {2012})}\BibitemShut {NoStop}%
\bibitem [{\citenamefont {G\"{u}nter}\ \emph {et~al.}(2013)\citenamefont
  {G\"{u}nter}, \citenamefont {Schempp}, \citenamefont {Robert-de
  Saint-Vincent}, \citenamefont {Gavryusev}, \citenamefont {Helmrich},
  \citenamefont {Hofmann}, \citenamefont {Whitlock},\ and\ \citenamefont
  {Weidem\"{u}ller}}]{gunter_observing_2013}%
  \BibitemOpen
  \bibfield  {author} {\bibinfo {author} {\bibfnamefont {G.}~\bibnamefont
  {G\"{u}nter}}, \bibinfo {author} {\bibfnamefont {H.}~\bibnamefont {Schempp}},
  \bibinfo {author} {\bibfnamefont {M.}~\bibnamefont {Robert-de
  Saint-Vincent}}, \bibinfo {author} {\bibfnamefont {V.}~\bibnamefont
  {Gavryusev}}, \bibinfo {author} {\bibfnamefont {S.}~\bibnamefont {Helmrich}},
  \bibinfo {author} {\bibfnamefont {C.~S.}\ \bibnamefont {Hofmann}}, \bibinfo
  {author} {\bibfnamefont {S.}~\bibnamefont {Whitlock}}, \ and\ \bibinfo
  {author} {\bibfnamefont {M.}~\bibnamefont {Weidem\"{u}ller}},\ }\href
  {\doibase 10.1126/science.1244843} {\bibfield  {journal} {\bibinfo  {journal}
  {Science}\ }\textbf {\bibinfo {volume} {342}},\ \bibinfo {pages} {954}
  (\bibinfo {year} {2013})}\BibitemShut {NoStop}%
\bibitem [{\citenamefont {Dudin}\ \emph
  {et~al.}(2012{\natexlab{a}})\citenamefont {Dudin}, \citenamefont {Bariani},\
  and\ \citenamefont {Kuzmich}}]{dudin_emergence_2012}%
  \BibitemOpen
  \bibfield  {author} {\bibinfo {author} {\bibfnamefont {Y.~O.}\ \bibnamefont
  {Dudin}}, \bibinfo {author} {\bibfnamefont {F.}~\bibnamefont {Bariani}}, \
  and\ \bibinfo {author} {\bibfnamefont {A.}~\bibnamefont {Kuzmich}},\ }\href
  {\doibase 10.1103/PhysRevLett.109.133602} {\bibfield  {journal} {\bibinfo
  {journal} {Phys. Rev. Lett.}\ }\textbf {\bibinfo {volume} {109}},\ \bibinfo
  {pages} {133602} (\bibinfo {year} {2012}{\natexlab{a}})}\BibitemShut
  {NoStop}%
\bibitem [{\citenamefont {Maxwell}\ \emph {et~al.}(2013)\citenamefont
  {Maxwell}, \citenamefont {Szwer}, \citenamefont {Paredes-Barato},
  \citenamefont {Busche}, \citenamefont {Pritchard}, \citenamefont {Gauguet},
  \citenamefont {Weatherill}, \citenamefont {Jones},\ and\ \citenamefont
  {Adams}}]{maxwell_storage_2013}%
  \BibitemOpen
  \bibfield  {author} {\bibinfo {author} {\bibfnamefont {D.}~\bibnamefont
  {Maxwell}}, \bibinfo {author} {\bibfnamefont {D.~J.}\ \bibnamefont {Szwer}},
  \bibinfo {author} {\bibfnamefont {D.}~\bibnamefont {Paredes-Barato}},
  \bibinfo {author} {\bibfnamefont {H.}~\bibnamefont {Busche}}, \bibinfo
  {author} {\bibfnamefont {J.~D.}\ \bibnamefont {Pritchard}}, \bibinfo {author}
  {\bibfnamefont {A.}~\bibnamefont {Gauguet}}, \bibinfo {author} {\bibfnamefont
  {K.~J.}\ \bibnamefont {Weatherill}}, \bibinfo {author} {\bibfnamefont
  {M.~P.~A.}\ \bibnamefont {Jones}}, \ and\ \bibinfo {author} {\bibfnamefont
  {C.~S.}\ \bibnamefont {Adams}},\ }\href {\doibase
  10.1103/PhysRevLett.110.103001} {\bibfield  {journal} {\bibinfo  {journal}
  {Phys. Rev. Lett.}\ }\textbf {\bibinfo {volume} {110}},\ \bibinfo {pages}
  {103001} (\bibinfo {year} {2013})}\BibitemShut {NoStop}%
\bibitem [{\citenamefont {Bettelli}\ \emph {et~al.}(2013)\citenamefont
  {Bettelli}, \citenamefont {Maxwell}, \citenamefont {Fernholz}, \citenamefont
  {Adams}, \citenamefont {Lesanovsky},\ and\ \citenamefont
  {Ates}}]{bettelli_exciton_2013}%
  \BibitemOpen
  \bibfield  {author} {\bibinfo {author} {\bibfnamefont {S.}~\bibnamefont
  {Bettelli}}, \bibinfo {author} {\bibfnamefont {D.}~\bibnamefont {Maxwell}},
  \bibinfo {author} {\bibfnamefont {T.}~\bibnamefont {Fernholz}}, \bibinfo
  {author} {\bibfnamefont {C.~S.}\ \bibnamefont {Adams}}, \bibinfo {author}
  {\bibfnamefont {I.}~\bibnamefont {Lesanovsky}}, \ and\ \bibinfo {author}
  {\bibfnamefont {C.}~\bibnamefont {Ates}},\ }\href {\doibase
  10.1103/PhysRevA.88.043436} {\bibfield  {journal} {\bibinfo  {journal} {Phys.
  Rev. A}\ }\textbf {\bibinfo {volume} {88}},\ \bibinfo {pages} {043436}
  (\bibinfo {year} {2013})}\BibitemShut {NoStop}%
\bibitem [{\citenamefont {Dudin}\ \emph
  {et~al.}(2012{\natexlab{b}})\citenamefont {Dudin}, \citenamefont {Li},
  \citenamefont {Bariani},\ and\ \citenamefont
  {Kuzmich}}]{dudin_observation_2012}%
  \BibitemOpen
  \bibfield  {author} {\bibinfo {author} {\bibfnamefont {Y.~O.}\ \bibnamefont
  {Dudin}}, \bibinfo {author} {\bibfnamefont {L.}~\bibnamefont {Li}}, \bibinfo
  {author} {\bibfnamefont {F.}~\bibnamefont {Bariani}}, \ and\ \bibinfo
  {author} {\bibfnamefont {A.}~\bibnamefont {Kuzmich}},\ }\href
  {http://dx.doi.org/10.1038/nphys2413} {\bibfield  {journal} {\bibinfo
  {journal} {Nat. Phys.}\ }\textbf {\bibinfo {volume} {8}},\ \bibinfo {pages}
  {790} (\bibinfo {year} {2012}{\natexlab{b}})}\BibitemShut {NoStop}%
\bibitem [{\citenamefont {Bariani}\ \emph {et~al.}(2012)\citenamefont
  {Bariani}, \citenamefont {Dudin}, \citenamefont {Kennedy},\ and\
  \citenamefont {Kuzmich}}]{bariani_dephasing_2012}%
  \BibitemOpen
  \bibfield  {author} {\bibinfo {author} {\bibfnamefont {F.}~\bibnamefont
  {Bariani}}, \bibinfo {author} {\bibfnamefont {Y.~O.}\ \bibnamefont {Dudin}},
  \bibinfo {author} {\bibfnamefont {T.~A.~B.}\ \bibnamefont {Kennedy}}, \ and\
  \bibinfo {author} {\bibfnamefont {A.}~\bibnamefont {Kuzmich}},\ }\href
  {\doibase 10.1103/PhysRevLett.108.030501} {\bibfield  {journal} {\bibinfo
  {journal} {Phys. Rev. Lett.}\ }\textbf {\bibinfo {volume} {108}},\ \bibinfo
  {pages} {030501} (\bibinfo {year} {2012})}\BibitemShut {NoStop}%
\bibitem [{\citenamefont {Petrosyan}\ and\ \citenamefont
  {M\o{}lmer}(2013)}]{petrosyan_2013}%
  \BibitemOpen
  \bibfield  {author} {\bibinfo {author} {\bibfnamefont {D.}~\bibnamefont
  {Petrosyan}}\ and\ \bibinfo {author} {\bibfnamefont {K.}~\bibnamefont
  {M\o{}lmer}},\ }\href {\doibase 10.1103/PhysRevA.87.033416} {\bibfield
  {journal} {\bibinfo  {journal} {Phys. Rev. A}\ }\textbf {\bibinfo {volume}
  {87}},\ \bibinfo {pages} {033416} (\bibinfo {year} {2013})}\BibitemShut
  {NoStop}%
\bibitem [{\citenamefont {Walker}\ and\ \citenamefont
  {Saffman}(2008)}]{walker_consequences_2008}%
  \BibitemOpen
  \bibfield  {author} {\bibinfo {author} {\bibfnamefont {T.~G.}\ \bibnamefont
  {Walker}}\ and\ \bibinfo {author} {\bibfnamefont {M.}~\bibnamefont
  {Saffman}},\ }\href {\doibase 10.1103/PhysRevA.77.032723} {\bibfield
  {journal} {\bibinfo  {journal} {Phys. Rev. A}\ }\textbf {\bibinfo {volume}
  {77}},\ \bibinfo {pages} {032723} (\bibinfo {year} {2008})}\BibitemShut
  {NoStop}%
\bibitem [{ric()}]{rick_thesis}%
  \BibitemOpen
  \href@noop {} {}\bibinfo {note} {Rick van Bijnen, PhD thesis, \textit{Quantum
  engineering with ultracold atoms}, 2013.}\BibitemShut {Stop}%
\bibitem [{met()}]{meta-stable}%
  \BibitemOpen
  \href@noop {} {}\bibinfo {note} {The two-atom steady-state obtained here is
  meta-stable state that decays with at a slow rate
  $\sim\Omega_p^2/\gamma$.}\BibitemShut {Stop}%
\bibitem [{\citenamefont {Boyd}(2008)}]{boyd_nonlinear_2008}%
  \BibitemOpen
  \bibfield  {author} {\bibinfo {author} {\bibfnamefont {R.~W.}\ \bibnamefont
  {Boyd}},\ }\href@noop {} {\emph {\bibinfo {title} {Nonlinear optics}}}\
  (\bibinfo  {publisher} {Academic Press},\ \bibinfo {address} {San Diego
  (California)},\ \bibinfo {year} {2008})\BibitemShut {NoStop}%
\bibitem [{\citenamefont {Vogt}\ \emph {et~al.}(2006)\citenamefont {Vogt},
  \citenamefont {Viteau}, \citenamefont {Zhao}, \citenamefont {Chotia},
  \citenamefont {Comparat},\ and\ \citenamefont {Pillet}}]{vogt_2006}%
  \BibitemOpen
  \bibfield  {author} {\bibinfo {author} {\bibfnamefont {T.}~\bibnamefont
  {Vogt}}, \bibinfo {author} {\bibfnamefont {M.}~\bibnamefont {Viteau}},
  \bibinfo {author} {\bibfnamefont {J.}~\bibnamefont {Zhao}}, \bibinfo {author}
  {\bibfnamefont {A.}~\bibnamefont {Chotia}}, \bibinfo {author} {\bibfnamefont
  {D.}~\bibnamefont {Comparat}}, \ and\ \bibinfo {author} {\bibfnamefont
  {P.}~\bibnamefont {Pillet}},\ }\href {\doibase 10.1103/PhysRevLett.97.083003}
  {\bibfield  {journal} {\bibinfo  {journal} {Phys. Rev. Lett.}\ }\textbf
  {\bibinfo {volume} {97}},\ \bibinfo {pages} {083003} (\bibinfo {year}
  {2006})}\BibitemShut {NoStop}%
\bibitem [{\citenamefont {Ryabtsev}\ \emph {et~al.}(2010)\citenamefont
  {Ryabtsev}, \citenamefont {Tretyakov}, \citenamefont {Beterov},\ and\
  \citenamefont {Entin}}]{ryabtsev_2010}%
  \BibitemOpen
  \bibfield  {author} {\bibinfo {author} {\bibfnamefont {I.~I.}\ \bibnamefont
  {Ryabtsev}}, \bibinfo {author} {\bibfnamefont {D.~B.}\ \bibnamefont
  {Tretyakov}}, \bibinfo {author} {\bibfnamefont {I.~I.}\ \bibnamefont
  {Beterov}}, \ and\ \bibinfo {author} {\bibfnamefont {V.~M.}\ \bibnamefont
  {Entin}},\ }\href {\doibase 10.1103/PhysRevLett.104.073003} {\bibfield
  {journal} {\bibinfo  {journal} {Phys. Rev. Lett.}\ }\textbf {\bibinfo
  {volume} {104}},\ \bibinfo {pages} {073003} (\bibinfo {year}
  {2010})}\BibitemShut {NoStop}%
\bibitem [{\citenamefont {Wu}\ \emph {et~al.}(2009)\citenamefont {Wu},
  \citenamefont {Artoni},\ and\ \citenamefont {La~Rocca}}]{RoccaPRL2009}%
  \BibitemOpen
  \bibfield  {author} {\bibinfo {author} {\bibfnamefont {J.-H.}\ \bibnamefont
  {Wu}}, \bibinfo {author} {\bibfnamefont {M.}~\bibnamefont {Artoni}}, \ and\
  \bibinfo {author} {\bibfnamefont {G.~C.}\ \bibnamefont {La~Rocca}},\ }\href
  {\doibase 10.1103/PhysRevLett.103.133601} {\bibfield  {journal} {\bibinfo
  {journal} {Phys. Rev. Lett.}\ }\textbf {\bibinfo {volume} {103}},\ \bibinfo
  {pages} {133601} (\bibinfo {year} {2009})}\BibitemShut {NoStop}%
\end{thebibliography}
\end{document}